\documentclass[11pt,graphicx]{article}
\usepackage{amssymb,amsmath,amsfonts}
\usepackage{graphicx}
\usepackage{graphics}
\usepackage{eepic,epsfig}
\begin{document}
\title{Vacuum polarization by a scalar field in de Sitter spacetime in the presence of global monopole}
\author{Eug\^enio R. Bezerra de Mello \thanks{E-mail: emello@fisica.ufpb.br}\\
Departamento de F\'{\i}sica-CCEN\\
Universidade Federal da Para\'{\i}ba\\
58.059-970, J. Pessoa, PB\\
C. Postal 5.008\\
Brazil}
\maketitle       
\begin{abstract} 
In this paper we analyse the vacuum polarization effects associated with massive scalar quantum fields in a higher-dimensional de Sitter space in the presence of a global monopole. Because this analysis has been developed in a pure de Sitter space, here we are mainly interested on the effects induced by the presence of the global monopole. So, in order to achieve this objective we calculate the corresponding Wightman function, which is expressed in an integral representation and explicitly depends on the parameters associated with the presence of the monopole and the cosmological constant. Admitting that the former is closed to the unity, which corresponds to a realistic value predicted by Grand Unified Theories, it is possible to express this function as the sum of two terms: the first one corresponds to the Wightman function on the bulk where the global monopole is absent, and the second one is a contribution induced by the presence of the monopole.  
\vspace{1pc}
\end{abstract}

\maketitle

\section{Introduction}
De Sitter (dS) spacetime is one of the most recurrent scenario to develop quantum field theories in curved space. One of the main reason is because it allows resolutions of relevant physical problem exactly; moreover the importance of corresponding theoretical investigation increased by the appearance of the inflationary cosmology scenario \cite{Linde}. For a great number of inflationary models, dS spacetime can be used to study relevant problems in standard cosmology. During an inflationary epoch, quantum fluctuations in the inflaton field introduce inhomogeneities which may affect the transition toward the true vacuum. These fluctuations play important role in the generation of cosmic structures from inflation. Recent astronomical observations data of high redshift supernovae, galaxy clusters and cosmic microwave background \cite{Ries}, indicate that at the present stage, the universe is accelerating and can be well approximated by a world with a positive cosmological constant.

It is well known that different types of topological defects may have been formed in the early universe after Planck time by the vacuum phase
transition \cite{Kibble,V-S}. Depending on the topology of the vacuum manifold these are domain walls, strings, monopoles and textures. Among
them, cosmic strings and monopoles seem to be the best candidate to be observed. Global monopoles are spherically symmetric topological stable gravitational defects which appear in the framework of grand unified theories. These objects could be produced as a consequence of spontaneous breakdown of a global $O(3)$ gauge symmetry \cite{BV}. The gravitational field produced by global monopoles may be approximated by a
solid angle deficit in the three-dimensional space.

Although topological defects have been first analyzed in four-dimensional spacetime, they have also been considered in the context of braneworld. In this scenario the defects live in a $n-$dimensions submanifold embedded in a $(4+n)-$ dimensional universe. The global monopole case, has been analysed in \cite{Ola,Cho} from higher-dimensional Einstein equation considering $n\geq3$.

The analysis of quantum effects associated with various spin fields on the background of dS spacetime, have been discussed by many authors; see for instance \cite{BD} to \cite{Mello} and references therein. Also for global monopole spacetime the corresponding analysis have been considered in \cite{M-L} to \cite{E-S}. Specifically in \cite{E} the calculation of the vacuum average of the field squared in a higher-dimensional global monopole has been developed. 

In this paper we shall investigate the quantum behavior associated with massive scalar fields in a higher-dimensional de Sitter space in the presence of a global monopole. The main physical motivation to develop this analysis is based on the fact that in the early stage of the Universe, during its inflationary phase described by a dS spacetime, vacuum fluctuations associated with matter fields do not depend on spacetime point. In this way the presence of defects during this stage will produce inhomogeneities for these quantities required to form cosmic structure in the Universe. Specifically in this paper, we shall see that the contributions on the vacuum expectation value (VEV) of scalar field squared and the energy-momentum tensor induced by the a global monopole depend on the radial distance to the defect, becoming much more relevant than the contribution due to the dS spacetime in the regions near the defect.

According to this objective two main analysis will be developed first: $(i)$ The construction of the Wightman function associated with a massive scalar field in the dS spacetime in the presence of a global monopole. As we shall see, although this function will be presented in a integral form containing an infinite products of Gegenbauer polynomials and associated Legendre functions, a more workable expression can be provided by considering that the parameter associated with the global monopole is close to the unity. $(ii)$ The calculation of  the vacuum expectation values of the field squared and the energy-momentum tensor, with the help of the obtained  Wightman function. These vacuum averages take into account the contributions due to the non-trivial topology of the manifold, associated with a solid angle deficit, and its curvature due cosmological constant and also the monopole. Because the VEVs of the field squared and the energy-momentum tensor in a pure dS space have been developed by many authors, here we are interested to calculate the contributions on these quantities induced by the global monopole.\footnote{Although global monopole produced in a phase transition before or during early stages of inflation are diluted by an enormous factor, monopole-antimonopole pairs can be continuously created during inflation by quantum-mechanical tunneling \cite{V-S}.}

This paper is organized as follows: in section \ref{WF} we present the gravitational background associated with the geometry under consideration and the solution of the Klein-Gordon equation admitting an arbitrary curvature coupling. Also we present the complete Wightman function. An approximated expression can be constructed considering realistic values for the parameter associated with the global monopole. For this case the Wightman function can be written as a sum of two contributions: the first one associated with a pure de Sitter manifold, and the second, a correction, consequence of the presence of the global monopole. In section \ref{Phi} we explicit calculate the contribution on the renormalized VEV of the field squared due to the presence of the global monopole, considering specific cases where the dimension of the manifold is $4$ and $5$. In section \ref{VEMT} we also calculate the contribution on the radial pressure, i.e., the $r-r$ component of the energy-momentum tensor due to the monopole, for a four dimensional manifold considering a conformal coupling. In section \ref{Conc} we present our conclusions and some important remarks about this paper. Finally we include in appendices \ref{Ap:A} and \ref{Ap:B} some technical details associated with important formulas used along the text.

\section{Wightman function}
\label{WF}
In this section we are interested to calculate the positive frequency Wightman function associated with a massive scalar field in the higher-dimensional de Sitter spacetime taking into account the presence of a global monopole. This quantity is important to calculate the corresponding vacuum polarization effects. To obtain this function we first calculate the complete set of eigenfunctions for the Klein-Gordon equation admitting an arbitrary curvature coupling. 

The geometry associated with a $D-$dimensional de Sitter (dS) spacetime in the presence of global monopole is given by the following line element:
\begin{eqnarray}
\label{dS0}
ds^2=dt^2-e^{2t/\alpha}\left(dr^2+\beta^2r^2d\Omega^2_{(p)}\right)=g_{\mu\nu}dx^\mu dx^\nu \ ,
\end{eqnarray}
where the coordinates, $x^\mu=(t, \ r, \ \theta_1, \ \theta_2, ...  ,\theta_{p-1}, \phi)$, are defined in the intervals $t\in(-\infty, \infty)$, $\theta_i\in [0, \pi]$ for $i= \ 1, \ 2 ..., \ p-1$, for $p\geq 2$, $\phi\in[0, \ 2\pi]$ and $r\geq 0$. The parameter $\beta$ which codifies the presence of the global monopole is smaller than unity. Moreover, the parameter $\alpha $ is related with the cosmological constant by the formula
\begin{equation}
\Lambda =\frac{(D-1)(D-2)}{2\alpha^{2}} \ ,
\label{Lambda}
\end{equation}
being $D=2+p$. For further analysis, in addition to the synchronous time coordinate $t$ we shall use the conformal time $\tau$\ defined according to
\begin{equation}
\tau =-\alpha e^{-t/\alpha }\ ,\ -\infty <\ \tau \ <\ 0\ .  \label{tau}
\end{equation}
In terms of this coordinate the above line element takes the form
\begin{eqnarray}
\label{dS1}
ds^2=(\alpha/\tau)^2\left(d\tau^2-dr^2-\beta^2r^2 d\Omega^2_{(p)}\right) \ . 
\end{eqnarray}
In this coordinate system the metric tensor is explicitly defined as shown below:
\begin{eqnarray}
\label{Metric}
g_{00}&=&(\alpha^2/\tau^2) \nonumber \\ 
g_{11}&=&-(\alpha^2/\tau^2) \nonumber \\ 
g_{22}&=&-(\beta^2\alpha^2r^2/\tau^2) \nonumber\\ 
g_{jj}&=&-(\beta^2\alpha^2r^2/\tau^2)\sin^2\theta_1\sin^2\theta_2 ...\sin^2\theta_{j-2} \ ,
\end{eqnarray}
for $3\leq j \leq p+1$, and $g_{\mu\nu}=0$ for $\mu\neq \nu$. This manifold corresponds to a point-like global monopole in de Sitter spacetime. The scalar curvature reads:
\begin{eqnarray}
	R=\frac{D(D-1)}{\alpha^2}+p(p-1)\frac{(1-\beta^2)\tau^2}{\alpha^2\beta^2r^2} \ .
\end{eqnarray}

The line element (\ref{dS0}) can also be written in static coordinates. For simplicity we consider the corresponding coordinate transformation in $D=4$
case with $p=2$. This transformation is given by the relations
\begin{eqnarray}
t &=&-t_s+\frac\alpha2\ln (1-r_s^2/\alpha^2) \nonumber\\
r&=&\frac{r_se^{t_s/\alpha}}{\sqrt{1-r_s^2/\alpha^2}} \nonumber\\
\theta &=&\theta \ {\rm and} \ \phi=\phi \ . 
\end{eqnarray}%
and the line element takes the form
\begin{equation}
ds^{2}=(1-r_{s}^{2}/\alpha ^{2})dt_{s}^{2}-\frac{dr_{s}^{2}}{1-r_{s}^{2}/\alpha ^{2}}-\beta^2r_{s}^{2}(d\theta ^{2}+\sin ^{2}\theta d\phi ^{2}) \ .
\label{dSstatic}
\end{equation}
Changing the radial coordinate $r_s\to r_s/\beta$, the above line element coincides with asymptotic expression found in \cite{Xin,Bruno}, which represents the leading order in gravitational coupling of the effect of a global monopole in de Sitter spacetime.

The field equation that governs the behavior of the field is the Klein-Gordon one considered below
\begin{equation}
(g^{\mu\nu}\nabla_\mu\nabla_\nu+m^{2}+\xi R)\Phi (x)=0 \ ,  \label{KG}
\end{equation}
where $\xi$ is an arbitrary curvature coupling constant. The complete set of solutions of this equation in the coordinate system defined by (\ref{dS1}) is:
\begin{equation}
\Phi_{\sigma }(x)=C_{\sigma}\frac{\eta^{(D-1)/2}}{r^{(p-1)/2}}H_{\mu }^{(1)}(\beta\omega\eta)J_{\nu_n}(\beta\omega r)Y(n, \ m_j; \ \theta_j, \ \phi) \ , \ \eta =\alpha e^{-t/\alpha } \ , \label{sol1}
\end{equation}%
where $H_{\mu }^{(2)}$ and $J_{\nu }$ represent the Hankel and Bessel functions respectively \cite{Grad}, of order 
\begin{eqnarray}
\label{order}
\mu&=&\frac12\sqrt{(D-1)^{2}-4\xi D(D-1)-4m^{2}\alpha ^{2}} \ , \nonumber\\
\nu_n&=&\frac1\beta\sqrt{\left(n+\frac{p-1}2\right)^2+p(p-1)(1-\beta^2)(\xi-\overline{\xi})} \ ,
\end{eqnarray}
with $\overline{\xi}=(p-1)/4p$. The function $Y(n, \ m_j)$ represents the hyperspherical harmonics of degree $n$ \cite{EMOT}. In (\ref{sol1}), $\sigma \equiv (\omega, \ n, \ m_j)$ is the set of quantum numbers, being $\omega\in \lbrack 0,\ \infty )$. The
coefficient $C_{\sigma }$ can be found by the orthonormalization condition
\begin{equation}
-i\int d^{D-1}x\sqrt{|g|}g^{00}[\Phi_{\sigma}(x)\partial_{\eta}\Phi_{\sigma^{\prime}}^{\ast}(x)-\Phi _{\sigma^{\prime}}^{\ast }(x)\partial_{\eta}\Phi
_{\sigma}^{\ast }(x)]=\delta _{\sigma ,\sigma ^{\prime }}\ ,
\label{normcond}
\end{equation}
where the integral is evaluated over the spatial hypersurface $\eta =\mathrm{const}$, and $\delta _{\sigma ,\sigma ^{\prime }}$ represents the
Kronecker-delta for discrete index and Dirac-delta function for continuous ones. This leads to the result
\begin{equation}
C_\sigma^{2}=\frac{\pi\omega e^{-i(\mu^*-\mu)\pi/2}}{4\alpha^{D-2}\beta^{p-2}N(m)} \ ,  \label{coef}
\end{equation}
for the normalization coefficient. The expression for $N(m)$ is given in \cite{EMOT} and will not be necessary for the following discussion.

We shall employ the mode-sum formula to calculate the positive frequency Wightman function:
\begin{equation}
G(x,x^{\prime })=\sum_{\sigma }\Phi_{\sigma }(x)\Phi _{\sigma }^{\ast}(x^{\prime})\ .  \label{Green}
\end{equation}
Substituting (\ref{sol1}), with respective coefficient (\ref{coef}), into (\ref{Green}) we obtain
\begin{eqnarray}
\label{Wight}
G(x,x^{\prime })&=&\frac{\pi e^{i(\mu^*-\mu)\pi/2}}{4\alpha^{D-2}\beta^{p-2}(p-1)\Omega_p}\frac{(\eta\eta')^{(D-1)/2}}{(rr')^{(p-1)/2}}\int_0^\infty d\omega\omega\sum_n(2n+p-1)\times\nonumber\\
&&C_n^{(p-1)/2}(\cos \gamma)J_{\nu_n}(\beta\omega r)J_{\nu_n}(\beta\omega r')H_\mu^{(1)}(\beta\omega\eta)(H_\mu^{(1)}(\beta\omega\eta'))^* \ ,
\end{eqnarray}
where we have used the addition  theorem for hyperspherical harmonics \cite{EMOT}:
\begin{eqnarray}
\label{YS}
\sum_{m_j}Y(n, \ m_j; \ \theta_j, \ \phi)Y^*(n, \ m_j; \ \theta_j', \ \phi')=N(n)\frac{(2n+p-1)}{(p-1)\Omega_p}C_n^{(p-1)/2}(\cos\gamma) \ ,
\end{eqnarray}
being $C_n^q$ the Gegenbauer polynomial of degree $n$ and order $q$. In (\ref{YS}), $\gamma$ is the angle defined between the two directions $(\theta_j, \ \phi)$ and $(\theta_j', \ \phi')$, and $\Omega_p=2\pi^{(p+1)/2}/\Gamma((p+1)/2)$, the solid angle associated with a hypersphere of unity radius. In appendix \ref{Ap:A} we show that (\ref{Wight}) can be written in a more workable form:
\begin{eqnarray}
\label{Wight1}
G(x,x^{\prime })&=&\frac1{(\alpha\beta)^p(p-1)\pi\Omega_p}\left(\frac{\eta\eta'}{rr'}\right)^{(D-1)/2}\int_0^\infty \ dy\frac{\cosh(2\mu y)}{\sinh(v)}\times\nonumber\\
&&\sum_n(2n+p-1)C_n^{(p-1)/2}(\cos\gamma)e^{-\nu_n v} \ ,
\end{eqnarray}
where the variable $v$ is defined by
\begin{equation}
	\cosh(v)=\frac1{2rr'}[r^2+r'^2-(\eta-\eta')^2+4\eta\eta'\sinh^2(y)] \ .
\end{equation}
Also in appendix \ref{Ap:A} it is shown that for $\beta=1$, the Wightman function can be expressed as \cite{Bell-Saha}:
\begin{eqnarray}
\label{G-dS}	G_{dS}(x,x')=\frac{\alpha^2}{2(2\pi\alpha^2)^{D/2}}\frac{\Gamma\left(\frac{D-1}2-\mu\right)\Gamma\left(\frac{D-1}2+\mu\right)}{(1-u^2)^{(D-2)/4}}P^{1-D/2}_{\mu-1/2}(u) \ ,
\end{eqnarray}
being $P^\nu_\mu$ the associated Legendre function of the first kind, and
\begin{equation}
u=-1+\frac{({\vec{x}}-{\vec{x}}')^2-(\eta-\eta')^2}{2\eta\eta'} \ .
\end{equation}

Having obtained the Wightman function (\ref{Wight1}) for an arbitrary value of $p$, our next steps will devoted to the calculation of renormalized vacuum expectation value of the field squared operator. Explicitly we shall develop this calculation considering $p=2$ and $p=3$.

\section{The computation of $\langle \Phi ^{2}\rangle$}
\label{Phi}
The calculation of the vacuum expectation value (VEV) of the square of the field operator is obtained by computing the Wightman function at the coincidence limit:
\begin{equation}
\langle\Phi^2(x)\rangle=\lim_{x'\to x}G(x,x') \ .
\end{equation}
However this procedure provides a divergent result. In order to obtain a finite and well defined result we must apply in this calculation some renormalization procedure. Here we shall adopt the point-splitting renormalization one. This procedure is based upon a divergence subtraction scheme in the coincidence limit of the Wightman function. This will be done in a manifest form, by subtracting from the Wightman function the singular part of the Hadamard expansion before applying the coincidence limit, as shown below:
\begin{equation}
\label{Phi1}
\langle\Phi^2(x)\rangle_{Ren.}=\lim_{x'\to x}\left[G(x,x')-G_{H}(x,x')\right] \ .
\end{equation}

Because in general it is not possible to provide a closed expression for the scalar Green function in a pure global monopole spacetime, many investigations about vacuum polarization effects, started with Loust\'o and Mazzitelli in \cite{M-L}, have been developed by using an approximated expression for the corresponding Green function. The reason resides in the non trivial dependence of the order of the Bessel function, $\nu_n$, with the angular quantum number $n$. This is explicitly shown in (\ref{order}) for this analysis. With the objective to provide a more workable Wightman function which allows us to develop the calculation of the VEV of the field squared, in appendix \ref{Ap:A} we present an approximated expression for (\ref{Wight1}), considering that the parameter $\beta$ is close to the unity.\footnote{In fact for a typical grand unified theories the parameter $\beta^2\approx 1-10^{-5}$.} By developing an expansion on the parameter $\Delta^2=1-\beta^2<<1$, for $\gamma=0$, up to the first order in this parameter, the Wightman function given in (\ref{Wight1}) can be expressed by:
\begin{eqnarray}
\label{Wight2}
G(x,x^{\prime })&=&\frac1{(2\alpha\beta)^p\pi\Omega_p}\left(\frac{\eta\eta'}{rr'}\right)^{(D-1)/2}\int_0^\infty \ dy\frac{\cosh(2\mu y)}{\sinh^{p+1}(v/2)}\times\nonumber\\
&&\left[1-\frac{pv\Delta^2}{2\sinh(v)}\left(1+4\xi\sinh^2(v/2)\right)\right] \ .
\end{eqnarray}
Moreover, expanding the factor $1/\beta^p$ up to the first power in $\Delta^2$, (\ref{Wight2}) can be expressed as the sum of two terms, the first one corresponds to the Wightman function in a pure dS spacetime, and the second term, corresponds to the contribution coming from the presence of the monopole, as shown below:
\begin{equation}
\label{dS-gm}
G(x,x')=G_{dS}(x,x')+G_{gm}(x,x') \ ,
\end{equation}
where
\begin{eqnarray}
\label{gm}
G_{gm}(x,x')&=&\frac{p\Delta^2}{2(2\alpha)^p\pi\Omega_p}\left(\frac{\eta\eta'}{rr'}\right)^{(D-1)/2}\int_0^\infty \ dy\frac{\cosh(2\mu y)}{\sinh^{p+1}(v/2)}\times\nonumber\\
&&\left[1-\frac{v}{\sinh(v)}\left(1+4\xi\sinh^2(v/2)\right)\right] \ .
\end{eqnarray}

As consequence of this procedure, the VEV of the field squared, (\ref{Phi1}), is given as the sum of two distinct contributions. The first one corresponds the value in a pure dS spacetime, and the second contribution due to the presence of the global monopole:
\begin{equation}
\langle\Phi^{2}(x)\rangle_{Ren} =\langle\Phi^{2}(x)\rangle_{dS}+\langle\Phi^{2}(x)\rangle_{gm} \ .  
\end{equation}
Due to the maximal symmetry of dS spacetime, $\langle\Phi^{2}(x)\rangle_{dS}$ does not depend on the spacetime point. Specifically $D=4$ this renormalized VEV is given by \cite{Cand75,Dowk76,Bunc78}
\begin{eqnarray}	\langle\Phi^2(x)\rangle_{dS}&=&\frac1{8\pi^2\alpha^2}\left\{\left(\frac{m^2\alpha^2}2+6\xi-1\right)\left[\Psi\left(\frac32+\nu\right)+\Psi\left(\frac32-\nu\right)
	-2\ln(m\alpha)\right]\right.\nonumber\\
	&-&\left.\frac{(6\xi-1)^2}{m^2\alpha^2}+\frac1{30m^2\alpha^2}-6\xi+\frac23\right\} \ ,
\end{eqnarray}
where $\Psi(x)$ is the logarithimic derivative of the gamma-function. 

The main objective of this paper is to calculate the contribution to the vacuum polarization due to the presence of the global monopole. This contribution is given by:
\begin{equation}
\label{Phi2}
\langle \Phi^{2}(x)\rangle_{gm}=\lim_{x'\to x}\left[G_{gm}(x,x')-{\tilde{G}}_{H}(x,x')\right] \ , 
\end{equation}
where ${\tilde{G}}_{H}(x,x')$ corresponds the expansion in the Hadamard function due only to the presence of the monopole.

In \cite{Chris1,Chris2}, Christensen has given a general expression for the Hadamard function for any arbitrary dimensional spacetime. In fact this function is given in a expansion form, expressed in terms of the square of the geodesic distance $2\sigma(x,x')$. The singular behavior of the Hadamard expansion crucially depends on the dimension of the spacetime; moreover for an even dimensional spacetime the expansion presents a logarithmic divergence. So because of this fact we shall develop, separately, the calculation of the renormalized vacuum  expectation value of the field squared, (\ref{Phi2}), for $D=4$ and $5$ in the following.

\subsection{Four dimensional spacetime}
\label{subsec:A}
For a four dimensional spacetime the Wightman function (\ref{Wight2}), reads:
\begin{eqnarray}
\label{W4}
G(x,x^{\prime })=\frac1{16\pi^2\alpha^2\beta^2}\left(\frac{\eta\eta'}{rr'}\right)^{3/2}\int_0^\infty \ dy\frac{\cosh(2\mu y)}{\sinh^3(v/2)} \left[1-\frac{v\Delta^2}{\sinh(v)}\left(1+4\xi\sinh^2(v/2)\right)\right] \ ,
\end{eqnarray}
being $\mu=\sqrt{9/4-12\xi-m^2\alpha^2}$. The above integral is divergent for $\mu\geq3/2$. 

In the calculation of (\ref{Phi2}), we may take first the coincidence limit of the conformal time, i.e., we assume $\eta=\eta'$. Doing the same procedure as we have done to express (\ref{dS-gm}), the global monopole contribution to the Wightman function, Eq. (\ref{gm}), reads:
\begin{eqnarray}
\label{gm1}
G_{gm}(x,x')=\frac1{16\pi^2\alpha^2}\frac{\Delta^2}{(\rho\rho')^{3/2}}\int_0^\infty \ dy\frac{\cosh(2\mu y)}{\sinh^3(v/2)}
\left[1-\frac{v}{\sinh(v)}\left(1+4\xi\sinh^2(v/2)\right)\right] \ ,
\end{eqnarray}
where
\begin{equation}
\cosh(v)=\frac{\rho^2+\rho'^2}{2\rho\rho'}+\frac2{\rho\rho'}\sinh^2(y) \ .
\end{equation}
In the above expressions we have introduced the dimensionless coordinate $\rho=r/\eta$, which corresponds to the proper distance from the monopole, $\alpha r/\eta$, measured in units of dS curvature radius, $\alpha$.

The presence of the global monopole modifies the curvature of the background manifold, consequently introducing new divergences at the radial coordinate coincidence limit, $r'=r$, in the Hadamard function. For a four dimensional spacetime, the singular expansion of the Hadamard functions reads \cite{Chris1}:
\begin{eqnarray}
G_H(x,x')=\frac1{8\pi^2\sigma(x,x')}+\frac1{16\pi^2}\left[m^2+\left(\xi-\frac16\right)R\right]\left[2\gamma+\ln\left(\frac{m^2_{DS}\sigma(x,x')}2\right)\right] \ ,
\end{eqnarray}
where $m_{DS}$ is equal to $m$ for massive scalar field, and equal to $\mu$, an arbitrary infrared cutoff energy scale, for a massless field. The curvature scalar $R$ corresponds to dS one, $R_{dS}=\frac{12}{\alpha^2}$, plus the correction due to the presence of the monopole, $R_{gm}=2\frac{(1-\beta^2)\eta^2}{\alpha^2\beta^2r^2}$. In our approximation we may use $R_{gm}\approx 2\frac{\Delta^2\eta^2}{\alpha^2r^2}$. As we can see the monopole introduced an extra logarithmic singularity in the Hadamard function. Consequently the term in the Hadamard function needed to renormalize (\ref{Phi2}) is:
\begin{equation}
{\tilde{G}}_{H}(x,x')=\frac{\Delta^2}{8\pi^2\alpha^2}\frac{1}{\rho^2}\left(\xi-\frac16\right)\left[2\gamma+\ln\left(\frac{m^2\sigma(x,x')}2\right)\right] \ .
\end{equation}

Substituting into the above expression, the radial one-half of the geodesic distance given by $\sigma(x,x')=\frac{\alpha^2(r-r')^2}{2\eta^2}$, we obtain
\begin{equation}
\label{H1}
{\tilde{G}}_{H}(x,x')=\frac{\Delta^2}{4\pi^2\alpha^2}\frac{1}{\rho^2}\left(\xi-\frac16\right)\ln\left(\frac{m\alpha(\rho-\rho')}2\right)+\frac{\Delta^2}{4\pi^2\alpha^2}\frac{\gamma}{\rho^2}\left(\xi-\frac16\right) \ .
\end{equation}
At this point it is convenient to express the logarithmic singular contribution in (\ref{H1}) by an integral form. This can be done by using the Legendre function of second kind $Q_0$, in an integral representation \cite{Grad}:
\begin{equation}
\label{Int.Repr}
Q_0(\cosh (u))=\frac1{\sqrt{2}}\int_u^\infty \frac{dy \ e^{-y/2}}{\sqrt{\cosh(y)-\cosh(u)}} \ ,
\end{equation}
being $\cosh(u)=\frac{\rho^2+\rho'^2}{2\rho\rho'}$. Doing this we find
\begin{eqnarray}
\label{H2}
{\tilde{G}}_{H}(x,x')&=&\frac{\Delta^2}{4\pi^2\alpha^2}\frac{1}{\rho^2}\left(\xi-\frac16\right)\ln\left(\frac{m\alpha(\rho+\rho')}2\right)+\frac{\Delta^2}{4\pi^2\alpha^2}\frac{\gamma}{\rho^2}\left(\xi-\frac16\right)\nonumber\\
&-&\frac{\Delta^2}{4\pi^2\alpha^2}\frac{1}{\rho^2}\left(\xi-\frac16\right)Q_0(\cosh(u)) \ .
\end{eqnarray}

\begin{figure}[tbph]
\begin{center}
\begin{tabular}{cc}
\epsfig{figure=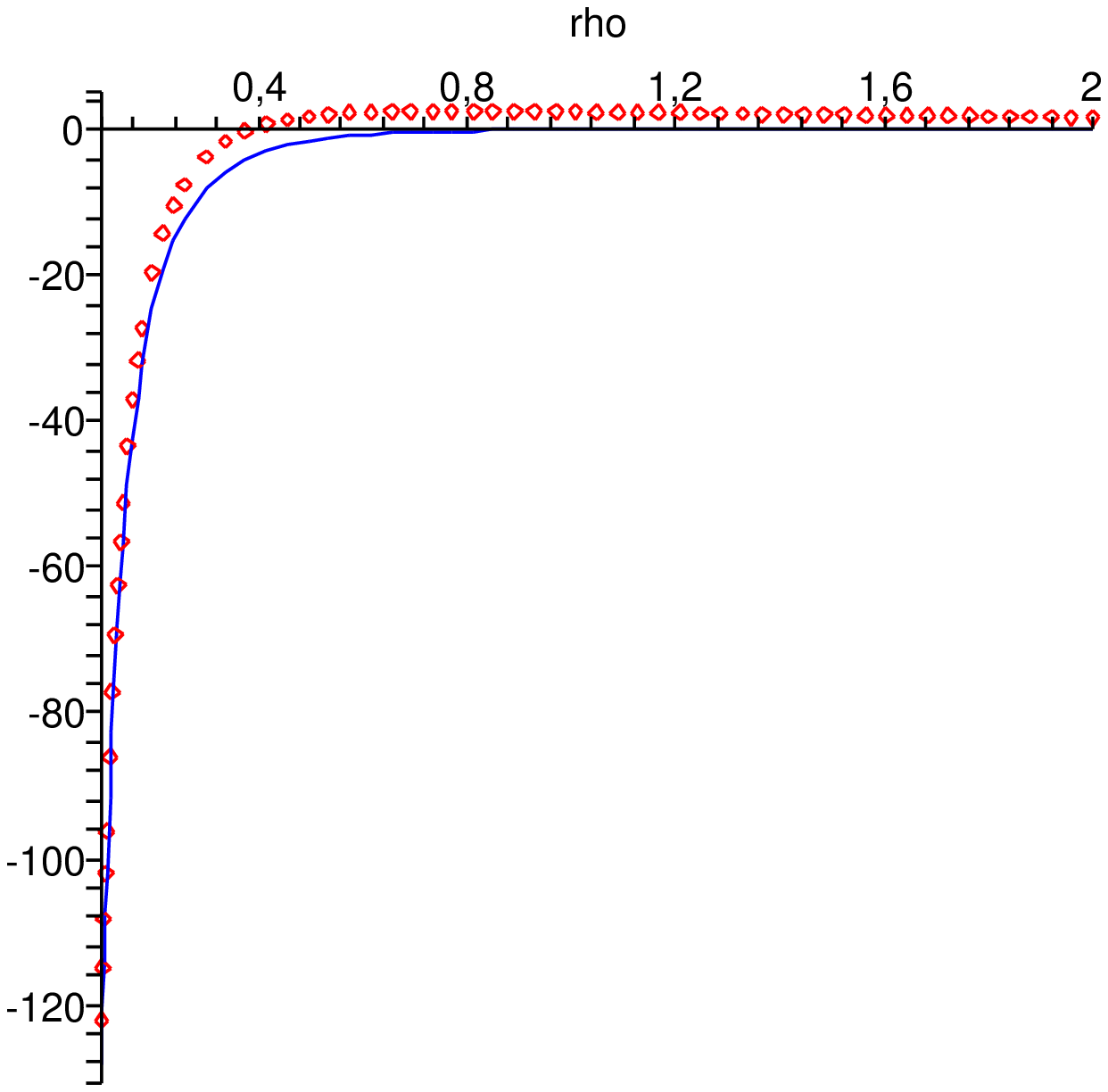, width=7.5cm, height=7.5cm,angle=0} & \quad 
\epsfig{figure=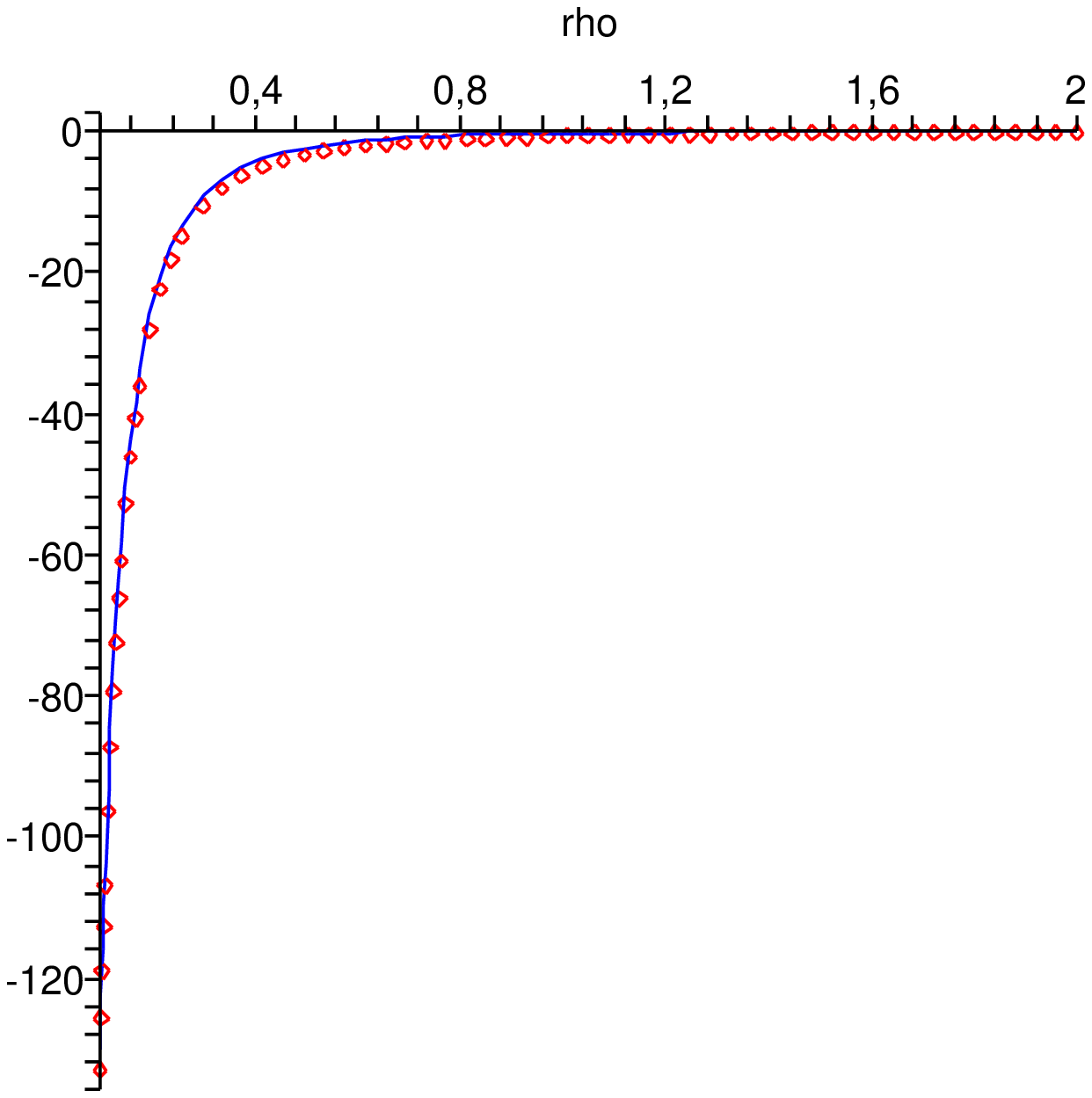, width=7.5cm, height=7.5cm,angle=0}
\end{tabular}
\end{center}
\caption{The behavior of the integral contribution induced by the global monopole in the VEV of the field squared, Eq. (\ref{PhiD4}), as a function of the ratio $\rho=r/\eta $ for a minimally coupled scalar field with $m\alpha =1, \ \sqrt{2}$ (left panel), and $m\alpha =\sqrt{14}/2, \ \sqrt{10}/2$ (right panel) in four-dimensional spacetime. For the first values, the behavior are represented by points, and for the seconds ones by solid line}
\label{fig1}
\end{figure}

Now substituting (\ref{gm1}) and (\ref{H2}) into (\ref{Phi2}), and after some intermediate steps, we obtain:
\begin{eqnarray}
\label{PhiD4}
\langle \Phi^{2}(x)\rangle_{gm}&=&\frac{\Delta^2}{16\pi^2\alpha^2}\int_0^\infty  dy  \left\{\frac{\cosh(2\mu y)}{\sinh^3(y)}\left[1- \frac{{\tilde{v}}}{2\sinh(y)\sqrt{\rho^2+\sinh^2(y)}}\left(\rho^2+4\xi\sinh^2(y)\right)\right]\right.\nonumber\\
&+&\left.\frac{2(\xi-1/6)}{\rho^2}\frac{e^{-y/2}}{\sinh(y/2)}\right\}-\frac{\Delta^2}{4\pi^2\alpha^2}\frac1{\rho^2}\left(\xi-\frac16\right)
\ln(m\alpha\rho)-\frac{\Delta^2}{4\pi^2\alpha^2}\frac\gamma{\rho^2}\left(\xi-\frac16\right) 
\end{eqnarray}
where
\begin{equation}
\label{v}
{\tilde{v}}=2 \ {\rm arcsinh}\left(\frac{\sinh(y)}{\rho}\right) \ .
\end{equation}
As we can see, the global monopole induced part in the VEV of the field squared, presents divergences at the monopole's position, given explicitly by the terms proportional to $\frac1{\rho^2}$ and $\frac{\ln(m\alpha\rho)}{\rho^2}$. Unfortunately it is not clear the behavior in (\ref{PhiD4}) given by the integral contribution. So, in order to exhibit this behavior, in figure \ref{fig1} we have plotted this contribution as function of the ratio $\rho=r/\eta$ (proper distance from the monopole measured in units of $\alpha $) for minimally coupled scalar field ($\xi =0$), with $m\alpha =1$, represented by points, and $m\alpha=\sqrt{2}$, by solid line, in the left panel; and with $m\alpha =\sqrt{14}/2$, by points, and $m\alpha=\sqrt{10}/2$, by solid line, in the right panel. For the left panel the correspondent values for $\mu$ are reals and for the right panel they are imaginaries. Also by this figure we can observe that for $\rho\to 0$, all the integrals become large and negative values, indicating a possible divergence. In fact analysing the integrand of (\ref{PhiD4}) for small values of $\rho$ we observe a divergence $\frac1{\rho^2}$, which vanishes for $\xi=1/6$, plus a logarithmic divergent term, $\ln(\rho)$. 

\subsection{Five dimensional spacetime}
For a five-dimensional spacetime, the contribution to the Wightman function due to the global monopole is:
\begin{eqnarray}
\label{gm:2}
G_{gm}(x,x')=\frac{3\Delta^2}{32\alpha^3\pi^3}\left(\frac{\eta\eta'}{rr'}\right)^{2}\int_0^\infty \ dy\frac{\cosh(2\mu y)}{\sinh^4(v/2)} \left[1-\frac{v}{\sinh(v)}\left(1+4\xi\sinh^2(v/2)\right)\right] \ .
\end{eqnarray}

As to the five-dimensional Hadamard function, in appendix \ref{Ap:B} it is given its singular expansion, as shown below:
\begin{eqnarray}
\label{GH5}
G_H(x,x')=\frac1{16\pi^2\sigma}\left[\frac1{\sqrt{2\sigma}}-\frac{m^2}2\sqrt{2\sigma}-\frac23m^3\sigma+\frac{a_1}2\sqrt{2\sigma}+a_1m\sigma\right] \ ,
\end{eqnarray}
where we have taken in this function, the coefficient $a_0(x,x')$ and the Van Vleck-Morette determinant, $\Delta(x,x')$, both equal to the unity in the obtained expression. The coefficient $a_1$ is equal to $(1/6-\xi)R$. Because we are interested to calculate the contribution to the VEV of the field squared induced by the global monopole, the Hadamard function used to renormalize (\ref{Phi2}) is:
\begin{equation}
\label{H3}
{\tilde{G}}_{H}(x,x')=\frac1{16\pi^2\sqrt{2\sigma}}\left(1/6-\xi\right)R_{gm}+\frac{m}{16\pi^2}\left(1/6-\xi\right)R_{gm} \ ,
\end{equation}
being $R_{gm}\approx \frac{6\Delta^2\eta^2}{\alpha^2r^2}$. Using the radial geodesic distance written in terms of the dimensionless coordinate, $\rho=r/\eta$, $\sigma=\frac{\alpha^2(\rho-\rho')^2}2$, and substituting this expression into (\ref{H3}) we obtain:
\begin{equation}
\label{H4}
{\tilde{G}}_{H}(x,x')=\frac{3\Delta^2}{8\pi^2\alpha^3}\frac1{\rho^2}\left(1/6-\xi\right)\frac1{(\rho-\rho')}+\frac{3\Delta^2}{8\pi^2}(1/6-\xi)\frac{m}{\alpha^2\rho^2} \ .
\end{equation}
As in the previous analysis, to calculate the contribution given by the monopole on the VEV of the field squared, it is convenient to express (\ref{H4}) in a integral form. Taking the coincidence limit of the conformal time first, $\eta'=\eta$, the representation adopted is:
\begin{equation}
\label{H5}
{\tilde{G}}_{H}(x,x')=\frac{3\Delta^2}{8\pi^2\alpha^3}\frac1{\rho^2}\left(1/6-\xi\right)\frac1{\pi\rho\rho'}\int_0^\infty \ dy \frac{\cosh(y)} {\sinh^2(v/2)}+\frac{3\Delta^2}{8\pi^2}(1/6-\xi)\frac{m}{\alpha^2\rho^2} \ .
\end{equation}

\begin{figure}[tbph]
\begin{center}
\begin{tabular}{cc}
\epsfig{figure=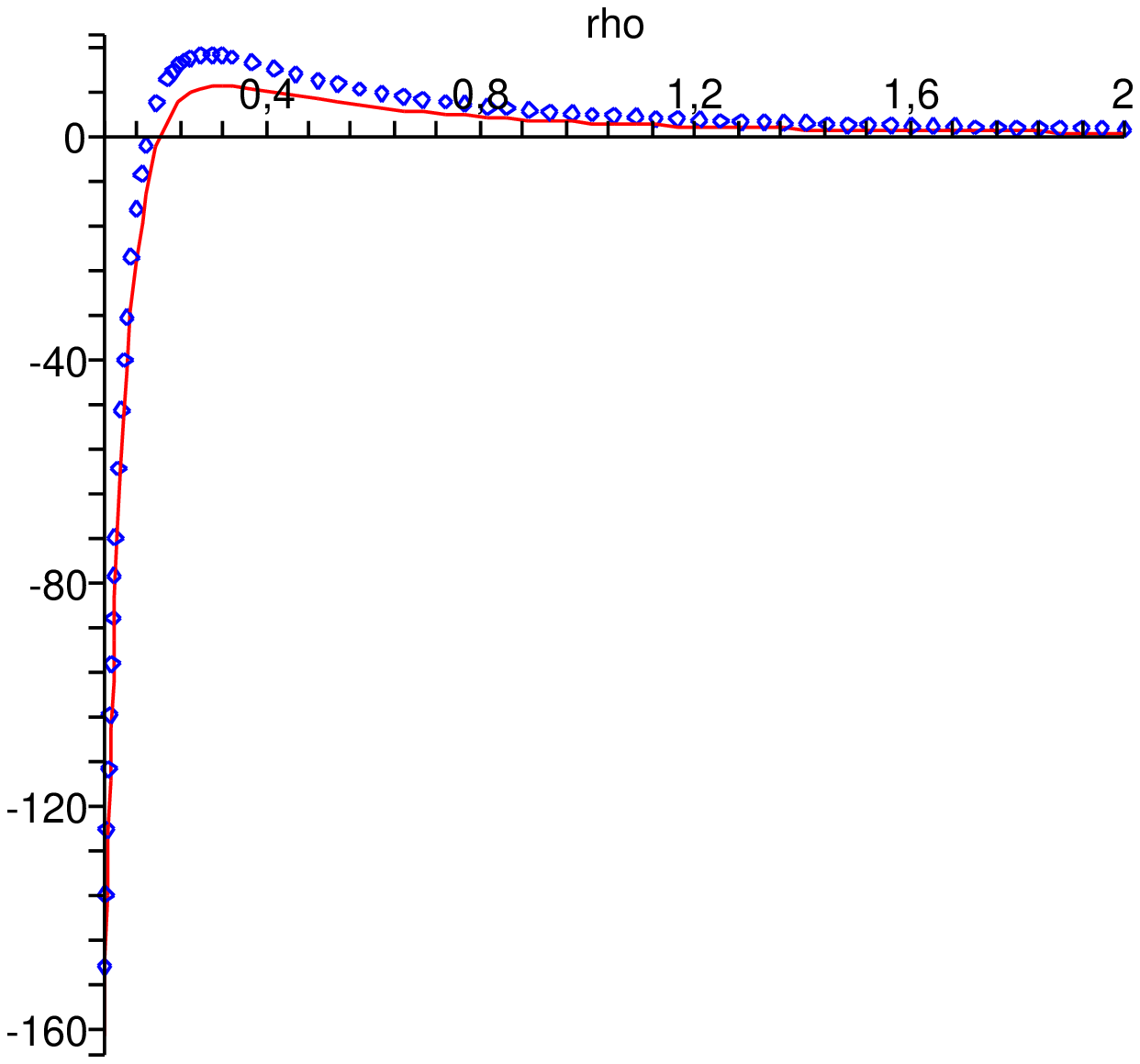, width=7.5cm, height=7.5cm,angle=0} & \quad 
\epsfig{figure=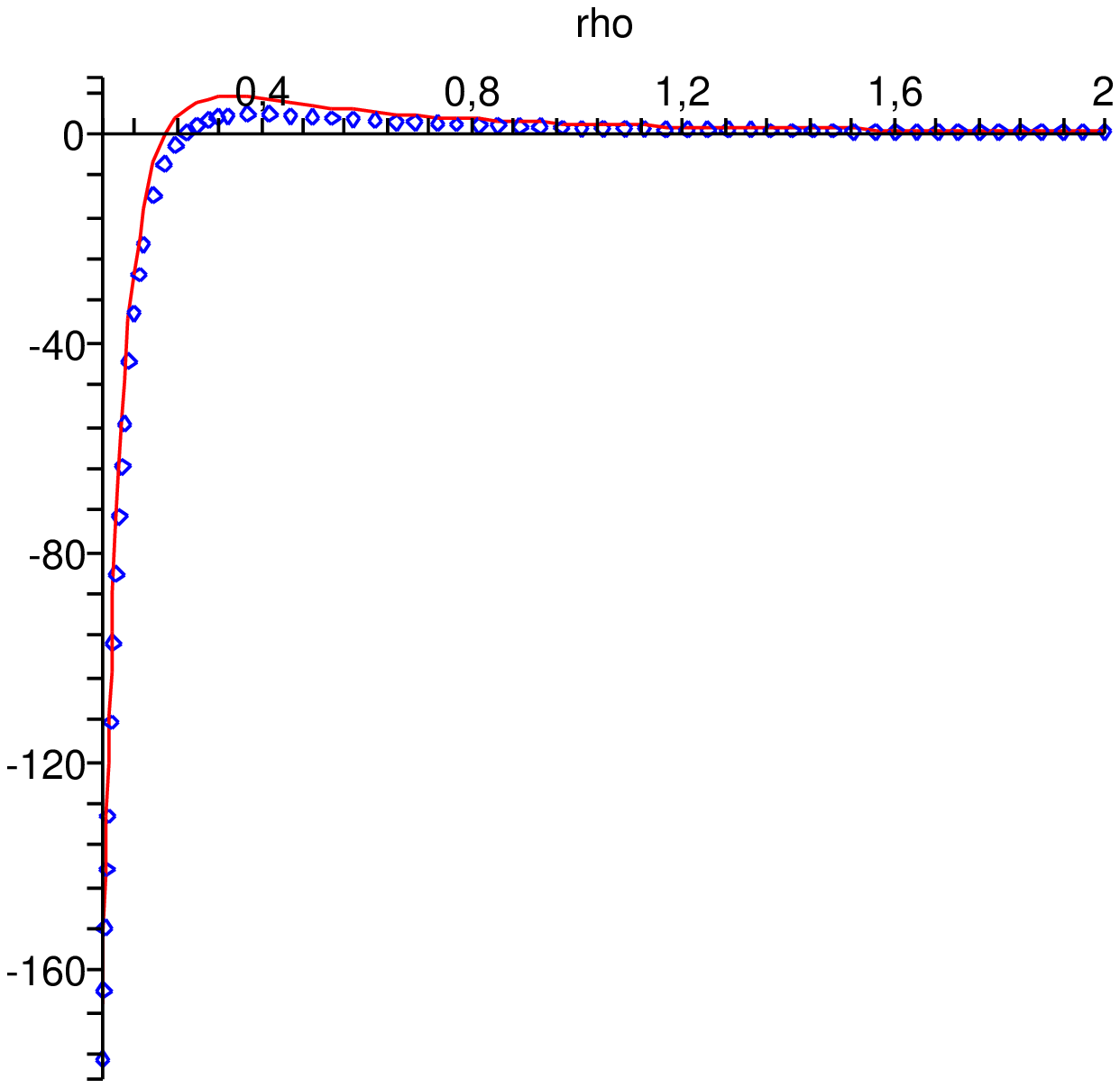, width=7.5cm, height=7.5cm,angle=0}
\end{tabular}
\end{center}
\caption{The behavior of the integral contribution induced by the global monopole in the VEV of the field squared as a function of the ratio $\rho=r/\eta$, for $\xi=0$ and $m\alpha =1, \ \sqrt{2}$ (left panel), and $m\alpha =\sqrt{14}/2, \ \sqrt{10}/2$ (right panel) in five-dimensional spacetime.}
\label{fig2}
\end{figure}

Substituting (\ref{gm:2}) and (\ref{H5}) into (\ref{Phi2}), we can express the contribution to the VEV of the field squared induced by the global monopole in an integral form, as shown below:
\begin{eqnarray}
\label{PhiD5}
\langle \Phi^{2}(x)\rangle_{gm}&=&\frac{3\Delta^2}{32\pi^3\alpha^3}\int_0^\infty  dy  \left\{\frac{\cosh(2\mu y)}{\sinh^4(y)}\left[1- \frac{{\tilde{v}}}{2\sinh(y)\sqrt{\rho^2+\sinh^2(y)}}\left(\rho^2+4\xi\sinh^2(y)\right)\right]\right.\nonumber\\
&-&\left.\frac4{\rho^2}(1/6-\xi)\frac{\cosh(y)}{\sinh^2(y)}\right\}-\frac{3\Delta^2}{8\pi^2}(1/6-\xi)\frac{m}{\alpha^2\rho^2} \ ,
\end{eqnarray}
with
\begin{equation}
{\tilde{v}}=2 \ {\rm arcsinh}\left(\frac{\sinh(y)}{\rho}\right) \ .
\end{equation}
As we can see this VEV is divergent at the monopole's position. This singular behavior is explicitly given by the term proportional to $\frac1{\rho^2}$. The behavior of the integral contribution in (\ref{PhiD5}) as function of the dimensionless variable $\rho$ is exhibited in figure \ref{fig2}. By these graphs, we can observe that this contribution becomes a negative large number for $\rho\to 0$. As in the previous analysis the values adopted for the coupling constant $\xi$ is the minimal one, and the values for $m\alpha$ are the same: $m\alpha =1$, represented by points, and $m\alpha=\sqrt{2}$, by solid line, in the left panel; and $m\alpha =\sqrt{14}/2$, by points, and $m\alpha=\sqrt{10}/2$, by solid line, in the right panel. We can see from figure \ref{fig2}, that there appear an oscillatory behavior in the VEV for small value of $\rho$.
\\

Before to finish this section we want to call attention for the fact that the presence of the global monopole in this background introduces position-dependent contributions to the VEV of the field square. These terms become dominant, when compared with the position-independent dS parts, in the region near the defect. 

\section{Vacuum expectation value of the energy-momentum tensor}
\label{VEMT} 
In this section we shall analyze the contribution to the VEV of the scalar energy-momentum tensor induced by the global monopole. This can be done explicitly by considering that the parameter $\beta$ is close to the unity. In this case the Wightman function can be expressed as the sum of two parts as shown in (\ref{dS-gm}), and consequently the VEV of the energy-momentum tensor can be written as shown below,
\begin{equation}
\langle T_{\mu\nu}\rangle =\langle T_{\mu\nu}\rangle_{dS}+\langle T_{\mu\nu}\rangle_{gm} \ ,  \label{Tmudec}
\end{equation}%
where $\langle T_{\mu\nu}\rangle_{dS}$ is the corresponding VEV in dS spacetime when the monopole is absent. This contribution, $\langle T_{\mu\nu} \rangle_{dS}$, does not depend on the spacetime point and has been well-investigated in literature \cite{Cand75,Dowk76,Bunc78}. It corresponds to a gravitational source of the cosmological constant type and, in combination with the initial cosmological constant $\Lambda $, given by (\ref{Lambda}), leads to the effective cosmological constant $\Lambda_{eff}=\Lambda +8\pi G\langle T_0^0\rangle_{dS}$, where $G$\ is the Newton gravitational
constant. The renormalized global monopole induced contribution can be obtained by using the formula \cite{Aram},
\begin{equation}
\langle T_{\mu \nu }\rangle_{gm}=\lim_{x'\to x}\partial_{\mu'}\partial_\nu[G_{gm}(x',x)-{\tilde{G}}_H(x',x)]+
\left[\left(\xi -1/4\right)g_{\mu \nu }\Box-\xi\nabla_\mu\nabla_\nu-\xi R_{\mu \nu }\right] \langle \Phi ^{2}\rangle_{gm} \ ,  \label{EMT}
\end{equation}
where $R_{\mu\nu}$ is the Ricci tensor in dS spacetime.

Before to embark in an explicit calculation of $\langle T_{\mu\nu}\rangle_{gm}$ we would like to say that, by the spherical symmetry which this problem contains, the only nonzero components of this tensor are: $\langle T_0^0\rangle_{gm}, \ \langle T_1^1\rangle_{gm}, \ \langle T_0^1\rangle_{gm} =\langle T^0_1\rangle_{gm} \ {\rm and} \ \langle T_2^2\rangle_{gm}=\langle T_3^3\rangle_{gm}$. Moreover, the complete tensor must be conserved,
\begin{equation}
\label{cons}
\nabla_\nu \langle T^\nu_\mu \rangle=0 \ ,
\end{equation}
and for the massless field conformally coupled, i.e., $\xi=\xi_c$, its trace satisfies the correct trace anomaly for an even dimensional spacetime:
\begin{equation}
\label{anom}
	\langle T^\nu_\mu \rangle_{ren}=\frac1{(4\pi)^{D/2}}a_{D/2} \ .
\end{equation}
Using both equations above, we can express all nonzero components of the energy-momentum tensor in terms of just one of them. Here, for simplicity we shall take the radial pressure, and develop the calculation for a four-dimensional spacetime considering $\xi=1/6$.

In order to calculate the radial pressure induced by the monopole, $\langle T_{11}\rangle_{gm}$, we should consider $\langle\Phi^2\rangle_{gm}$ given in (\ref{PhiD4}) with $\xi=1/6$. For this case $\mu=\frac12\sqrt{1-(2m\alpha)^2}$. The corresponding Wightman function needed is
\begin{eqnarray}
\label{gm2}
G_{gm}(x,x')=\frac{\Delta^2}{16\pi^2\alpha^2}\left(\frac{\eta\eta'}{rr'}\right)^{3/2}\int_0^\infty \ dy \ \frac{\cosh(2\mu y)}{\sinh^3(v/2)}
\left[1-\frac{v}{\sinh(v)}\left(1+\frac{2\sinh^2(v/2)}3\right)\right] \ ,
\end{eqnarray}
where
\begin{equation}
\cosh(v)=\frac{r^2+r'^2}{2rr'}-\frac{(\eta-\eta')^2}{2rr'}+\frac{2\eta\eta'}{rr'}\sinh^2(y) \ .
\end{equation}
Because we shall calculate the radial pressure only, we may take $\eta'=\eta$ in the above expressions.

The Hadamard function needed, should contains all terms which survive after taking the second derivative in the radial coordinate and the coincidence limit. This function reads \cite{Chris1,Chris2}:
\begin{eqnarray}
G_H(x,x')&=&\frac1{8\pi^2\sigma}+\frac1{8\pi^2}L(m^2-a_1)+\frac{m^2\sigma \ L}{16\pi^2}\left(\frac{m^2}2-a_1\right)+\frac{m^2 \ \sigma} {16\pi^2}\left(a_1- \frac{5m^2}{8}\right)\nonumber\\
&+&\frac{a_2 \ \sigma}{16\pi^2}\left(L-\frac12\right) \ ,
\end{eqnarray} 
where $L=\gamma+\frac12\ln\left(\frac{m^2\sigma}{2}\right)$, being $\gamma$ the Euler's constant. In \cite{Chris1} is given general expressions for the coefficients $a_1 \ {\rm and} \ a_2$. Because we are using $\xi=1/6$, $a_1=0$; moreover, the coefficient $a_2$ presents contributions coming from the pure dS space, plus from the global monopole. The contribution from the global monopole alone to the Hadamard function provides:
\begin{eqnarray}
\label{Hada1}
{\tilde{G}}_H(x,x')=\frac{{\tilde{a}}_2 \ \sigma}{16\pi^2}\left(L-\frac12\right) \ ,
\end{eqnarray} 
with ${\tilde{a}_2}=\frac{\Delta^2\eta^4}{45\alpha^4r^4}+{\rm O}(\Delta^4)$. Because we are considering $\eta=\eta'$ and also used the coincidence limit for the angular variables, the geodesic distance reads, $\sigma=\frac{\alpha^2(r-r')^2}{2\eta^2}$.

The next step is to take the second derivative on radial coordinates $r$ and $r'$ of (\ref{gm2}) and (\ref{Hada1}) and the coincidence limit, $r'=r$. As to the Wightman function, after some intermediate steps, we found the following expression:
\begin{eqnarray}
\label{53}
	\partial_r\partial_{r'}G_{gm}(x,x')&\longrightarrow &\frac{\Delta^2}{384\pi^2\alpha^2}\frac{\eta^2}{r^5}\int_0^\infty \ \frac{\cosh(2\mu y)}{\cosh^5(v/2)\sinh^6(v/2)}\left\{\sinh(v)[12+26\sinh^2(v/2)\right.\nonumber\\
&+&17\sinh^4(v/2)]-v[12+34\sinh^2(v/2)+33\sinh^4(v/2)\nonumber\\
&+&\left.14\sinh^6(v/2)]\right\} \ .
\end{eqnarray}
The behavior of the above integrand for small value of $y$ is:
\begin{equation}
\label{Limit}
\frac{\Delta^2}{720\pi^2\alpha^2}\frac{\eta^2}{r^4}\frac1y+O(y) \ .
\end{equation}
Consequently (\ref{53})  presents a logarithmic divergence.

As to the Hadamard function, we obtain:
\begin{equation}
\label{55}	\partial_r\partial_{r'}{\tilde{G}}_{H}(x,x')\longrightarrow-\frac{\Delta^2}{720\pi^2\alpha^2}\frac{\eta^2}{r^4}(1+\gamma)-\frac{\Delta^2}{720\pi^2\alpha^2}\frac{\eta^2}{r^4}\ln\left(\frac{m\alpha(r-r')}{2\eta}\right)+O((r-r')) \ .
\end{equation}
Using the same procedure adopted in subsection \ref{subsec:A} we express the logarithmic part in terms of the Legendre function $Q_0$:
\begin{eqnarray}
\ln\left(\frac{m\alpha(r-r')}{2\eta}\right)=\ln\left(\frac{m\alpha(\rho-\rho')}2\right)=\ln\left(\frac{m\alpha(\rho+\rho')}2\right)-Q_0(\cosh(u)) \ ,
\end{eqnarray}
being $\cosh(u)=\frac{\rho^2+\rho'^2}{2\rho\rho'}$. Using the integral representation for $Q_0$ given in (\ref{Int.Repr}), we observe that the corresponding integrand in (\ref{53}) presents the same divergent behavior as shown in (\ref{Limit}), for small variable $y$, when we take the coincidence limit, $\rho'\to\rho$. This makes that the contribution for the first term in (\ref{EMT}), for $\mu=r \ {\rm and} \ \mu'=r'$, becomes finite. Below we write down this final result:
\begin{eqnarray}
\label{part1}
&&\partial_r\partial_{r'}[G_{gm}(x,x')-{\tilde{G}}_H(x,x')]=\frac{\Delta^2}{720\pi^2\alpha^2}\frac{\eta^2}{r^4}\int_0^\infty \ \left\{\frac{15}{8\rho}\frac{\cosh(2\mu y)}{\cosh^5(v/2)\sinh^6(v/2)}\times\right.\nonumber\\
&&\left[\sinh(v)(12+26\sinh^2(v/2)+17\sinh^4(v/2))-v(12+34\sinh^2(v/2)+33\sinh^4(v/2)\right.\nonumber\\
&&+\left.\left.14\sinh^6(v/2))\right]-\frac{e^{-y/2}}{2\sinh(y/2)}\right\}+\frac{\Delta^2}{720\pi^2\alpha^2}\frac{\eta^2}{r^4}\ln(m\alpha\rho) +\frac{\Delta^2}{720\pi^2\alpha^2}\frac{\eta^2}{r^4}(1+\gamma) \ .
\end{eqnarray}
The integrand of the above expression goes to $1/2+O(y)$, for small value of $y$, and goes to zero for large value of this variable.

As to the second contribution to the radial pressure given by (\ref{EMT}), after a long calculation we obtain:
\begin{eqnarray}
\label{part2}
K_{rr}\langle\Phi^2\rangle_{gm}&=&-\frac{\Delta^2}{72\pi^2\alpha^2}\frac{\eta^3}{r^5}\int_0^\infty \ dy \ \frac{\cosh(2\mu y)} {\cosh(v/2)\sinh^4(v)}\{v[9-3\rho^2-\cosh^2(v/2)(23\nonumber\\
&-&20\cosh^2(v/2))+\rho^2\cosh^2(v/2)(13-34\cosh^2(v/2)-12\cosh^4(v/2))\nonumber\\
&+&\sinh(v)[9-3\rho^2-\cosh^2(v/2)(17-2\cosh^2(v/2))+\rho^2\cosh^2(v/2)(11\nonumber\\ 
&+&28\cosh^2(v/2))]\} \ .
\end{eqnarray}
Where we have represented the differential operator acting on the field squared in (\ref{EMT}) by $K_{rr}$. The integrand in the above expression behaves as $\frac{\Delta^2(26-\rho^2)}{1080\eta^2\pi^2\alpha^2\rho^6}y$ for small value of $y$, and goes to zero for $y\to\infty$.

From (\ref{part1}) and (\ref{part2}), we can see that both terms diverge for small distance from the global monopole. The complete behavior of both terms can only be provided numerically.

Here also, we can see that the contributions associated with he global monopole on the VEV of the energy-momentum tensor, becomes dominant for points near the defect.

\section{Conclusion}
\label{Conc}

In this paper we investigate the vacuum polarization effects associated with a scalar quantum field in a higher dimensional dS space in the presence of a global monopole. Two explicit calculations were developed: the VEV of the field squared and the $r-r$ component of the energy-momentum tensor. In order to do that, we have constructed the complete Wightman function corresponding to the physical situation under investigation. We presented it in an integral form which contains infinite sum of products of Gegenbauer polynomials by exponential functions. Unfortunately, due to the non-trivial dependence on  the angular quantum number with the parameter $\beta$, which codify the presence of the global monopole, it is not possible to provide a closed expression for this function. In order to continue with the investigation, we decided to consider an alternative procedure: we assumed that $\beta$ is closed to the unity, and developed a series expansion in the integrand of  the Wightman function in terms of the parameter $\Delta^2=1-\beta^2$, which measure the modification on the dS space caused by the monopole. Adopted this procedure we were able to express the Wightman function in two parts: the first one corresponding to the Wightman function on pure dS space, and the second one a contribution coming from the global monopole. This approach allows us to express the VEV of the field squared and the energy-momentum tensor, in terms of two distinct contributions in agreement to the obtained expression for the Wightman function. Because in the literature there are many paper related with the investigation of vacuum polarization effects in a pure dS space, here we focus our attention only in the global monopole induced contributions for both physical quantities. As to the VEV of the field squared, we have investigated the cases of space with four a five dimensions. These VEVs depend on the radial and time coordinate in the combination $\rho=r/\eta$, which is the proper distance from the monopole measured in units of the dS curvature. We have shown that the induced contributions become relevant for small values of $\rho$. In fact for these regions the induced contributions become larger than the contributions coming from the dS space, which by its turn does not depend on the spacetime point; on the other hand for points far from the monopole, or small conformal time, the situation becomes the opposite: the contributions form the dS space supplant the global monopole's ones. Although the complete behavior for the contributions induced by the monopole could not be presented in terms of closed expressions, we have provided numerical behaviors for the corresponding analysis. The latter were exhibited in figures \ref{fig1} and \ref{fig2}, for four and five-dimensional spacetimes, respectively. The next analysis developed was related with the vacuum polarization of the energy-momentum tensor. This investigation is much more complicated than the previous ones. There are more terms involved; so in order to provide a concrete result, we investigated only the $r-r$ component of this tensor, for a four-dimensional space, considering the conformal coupling. As far as we know this was the first time that the explicit calculation of the VEV of the energy-momentum tensor associated with quantum scalar field has been developed in a manifold involving a global monopole. Moreover, although we have not provided numerical analysis for this investigation, by resulted obtained for both contributions for the radial pressure, given in (\ref{part1}) and (\ref{part2}), we could observe that both are divergent for points near the monopole, and decrease as $r$ goes to infinity. We would like to finish this discussion by saying that the problem considered here is also of particular interest as an example where gravitational and topological polarizations of the vacuum are presented separately.

The analysis of global monopole evolution in a expanding universe has been developed in \cite{DBennett}. By numerical simulations the authors observed that as the Universe expands many annihilations  process between monopole and anti-monopole take place, and it is expected the order of $1$ monopole or anti-monopole per horizon volume.

The investigation of the vacuum polarization effects by a scalar massive field in higher-dimensional de Sitter spacetime and in presence of a cosmic string, have been developed in \cite{Mello1}. There it was emphasized that the presence of this linear topological defect does not introduce additional curvature. All the divergences presented in the evaluation of corresponding Wightman function at the coincidence limit, are contained in the pure dS spacetime part. The string induced part is finite for points outside the defect. Moreover, the analysis of VEVs associated with scalar and fermionic quantum fields in the presence of composite topological defects have been presented in \cite{Saha1} and \cite{Saha2}, respectively. In these papers a global monopole
and cosmic string have been considered in a higher-dimensional spacetime. There, by using similar procedure as presented here, we calculated the contribution induced by the cosmic string on the corresponding vacuum averages.

\section*{Acknowledgments}
E.R.B.M. thanks Conselho Nacional de Desenvolvimento Cient\'{\i}fico e Tecnol\'{o}gico (CNPq) for partial financial support, FAPESQ-PB/CNPq (PRONEX) and
FAPES-ES/CNPq (PRONEX). 

\appendix

\section{Integral representation for the Wightman function}
\label{Ap:A}
In order to transform the Wightman function given by (\ref{Wight}) in the expression (\ref{Wight1}), we present first the product of the Hankel functions in terms of the MacDonald functions,
\begin{equation}
e^{i(\mu -\mu ^{\ast })\pi /2}H_{\mu }^{(1)}(\lambda \eta )[H_{\mu}^{(1)}(\lambda \eta ^{\prime })]^{\ast }=\frac{4}{\pi ^{2}}K_{\mu}(-i\lambda \eta )K_{\mu }(i\lambda \eta ^{\prime }) \ ,  \label{HankProd}
\end{equation}
and use the formula \cite{Wats44}
\begin{equation}
K_{\mu}(a)K_{\mu }(b)=\frac{1}{2}\int_{0}^{\infty }\frac{dx}x \ e^{-1/2[x+(a^2+b^2)/x]}\int_{-\infty}^{+\infty }dy \ e^{-[2\mu y+ab\cosh (2y)/x]} \ .
\label{MacProd}
\end{equation}
As a result, the expression for the Wightman function is presented in the form
\begin{eqnarray}
G(x,x')&=&\frac{\beta^{2-p}}{2\pi\alpha^{D-2}(p-1)\Omega_p}\frac{(\eta\eta')^{(D-1)/2}}{(rr')^{(p-1)/2}}\sum_{n=0}^\infty (2n+p-1)C_n^{(p-1)/2}(\cos\gamma)\times \nonumber\\
&&\int _{-\infty}^\infty dy \ e^{-2\mu y}\int_0^\infty\frac{dx}x \ e^{-\frac{\vartheta x}2}\int_0^\infty d\omega\omega J_{\nu_n}(\omega\beta r)J_{\nu_n}(\omega\beta r') e^{-\frac{\beta^2\omega^2}{2x}}\ ,
\end{eqnarray}
where we have defined $\vartheta=2\eta\eta'\cosh(2y)-\eta^2-\eta'^2$. The function $C_n^q$ represents the Gegenbauser polynomials. The integration over the variable $\omega$ can be done with the help of \cite{Grad}, reproducing
\begin{eqnarray}
\label{WH}
G(x,x')&=&\frac1{2\pi\beta^p\alpha^{D-2}(p-1)\Omega_p}\frac{(\eta\eta')^{(D-1)/2}}{(rr')^{(p-1)/2}}\sum_{n=0}^\infty (2n+p-1)C_n^{(p-1)/2}(\cos\gamma)\times \nonumber\\
&&\int _{-\infty}^\infty dy \ e^{-2\mu y}\int_0^\infty{dx}\ e^{-\frac{x}2(\vartheta+r^2+r'^2)} I_{\nu_n}(rr'x) \ .
\end{eqnarray}
Being $I_\nu$ the modified Bessel function. Changing $x\to 2x$, and defining a new variable $z=e^{-2y}$, after some steps we obtain:
\begin{eqnarray}
G(x,x')&=&\frac1{2\pi\beta^p\alpha^{D-2}(p-1)\Omega_p}\frac{(\eta\eta')^{(D-1)/2}}{(rr')^{(p-1)/2}}\sum_{n=0}^\infty (2n+p-1)C_n^{(p-1)/2}(\cos\gamma)\times \nonumber\\
&&\int_0^\infty{dx}\int_0^\infty \frac{dz}{z^{\mu+1}}\ e^{-\eta\eta'x(z+1/z)} \ e^{-x(r^2+r'^2-\eta^2-\eta'^2)} I_{\nu_n}(rr'x) \ .
\end{eqnarray}
The integral over the variable $x$ can also be done. The result is given in terms of the Legendre function of the second kind, $Q_\nu$:
\begin{eqnarray}
G(x,x')&=&-i\frac{\alpha^{2-D}}{(2\pi)^{3/2}\beta^p(p-1)\Omega_p}\left(\frac{\eta\eta'}{rr'}\right)^{(D-1)/2}\sum_{n=0}^\infty (2n+p-1)C_n^{(p-1)/2}(\cos\gamma)\times \nonumber\\
&&\int_0^\infty \frac{dz}{z^{\mu+1}}\frac1{\sqrt{\sinh(v)}}Q_{\nu_n-1/2}^{1/2}(\cosh(v)) \ ,
\end{eqnarray}
where
\begin{equation}
\label{Cosv}
\cosh(v)=\frac1{2rr'}(r^2+r'^2-\eta^2-\eta'^2+2\eta\eta'\cosh(2y)) \ .
\end{equation}
Finally defining $z=e^{2y}$, using for the Legendre function the representation $Q_{\nu-1/2}^{1/2}=i\sqrt{\frac\pi{2\sinh(v)}}e^{-\nu v}$, and developing some intermediate steps, we find (\ref{Wight1}):
\begin{eqnarray}
\label{Wfunc}
G(x,x')&=&\frac1{(\alpha\beta)^p(p-1)\pi\Omega_p}\left(\frac{\eta\eta'}{rr'}\right)^{(D-1)/2}\int_0^\infty \ dy\frac{\cosh(2\mu y)} {\sinh(v)}\times\nonumber\\
&&\sum_{n=0}^\infty (2n+p-1)C_n^{(p-1)/2}(\cos\gamma)e^{-\nu_n v} \ .
\end{eqnarray}

Alternatively, by using  the integral representation for the Bessel function below,
\begin{equation}
\int_{-\infty }^{+\infty }dy\  e^{-2\mu y-2\eta \eta ^{\prime}\cosh (2y)u}=K_{\mu }(2\eta\eta^{\prime }u) \ ,
\end{equation}
we also can write, from (\ref{WH}), the Wightman function as shown below: 
\begin{eqnarray}
\label{Wfunc1}
G(x,x')&=&\frac{\alpha^{2-D}}{\beta^p(p-1)\pi\Omega_p}\frac{(\eta\eta')^{(D-1)/2}}{(rr')^{(p-1)/2}}\sum_{n=0}^\infty (2n+p-1)C_n^{(p-1)/2}(\cos\gamma)\times \nonumber\\
&&\int_0^\infty \ dx K_{\mu}(2\eta\eta'x)I_{\nu_n}(2rr'x)e^{-x(r^2+r'^2-\eta^2-\eta'^2)} \ .
\end{eqnarray}
This expression is important because taking $\beta=1$, we get $\nu_n=n+\frac{p-1}2$ and by using the relation below \cite{Prud} 
\begin{equation}
\sum_{n=0}^\infty(2n+p-1)C_n^{(p-1)/2}(\cos\gamma)I_{n+\frac{p-1}2}(z)=\frac2{\Gamma(\frac{p-1}2)}\left(\frac z2\right)^{(p-1)/2}e^{z\cos\gamma} \ ,
\end{equation}
we can integrate over the variable $x$ in (\ref{Wfunc1}), and obtain the Wightman function in de Sitter spacetime given in (\ref{G-dS}).

Returning to previous analysis, due to the non-trivial dependence of $\nu_n$ with $n$, see (\ref{order}), it is not possible to develop the summation on the quantum number $n$ in (\ref{Wfunc}), and to obtain a closed expression for the Wightman function. However, an approximated expression can be provided. Assuming that the parameter $\beta$ is close to the unity, we can develop a series expansion in $\Delta^2=1-\beta^2$, which is much smaller than unity. Moreover, because we want to take the coincidence limit in the Wightman function, we may take first $\gamma=0$. The value of the Gegenbauser polynomials when it argument is equal to unity is \cite{EMOT}:
\begin{equation}
C_n^{(p-1)/2}(1)=\frac{(n+p-2)!}{n!(p-2)!} \ .
\end{equation}
The approximated expression for $\nu_n$, up to the first power in $\Delta^2$, is
\begin{equation}
\nu_n\approx \left(n+\frac{p-1}2\right)(1+\Delta^2/2)+\frac{p(p-1)(\xi-\overline{\xi})}{2n+p-1}\Delta^2+O(\Delta^4) \ .
\end{equation}
We also need to develop the summation
\begin{equation}
S=\sum_{n=0}^\infty(2n+p-1)\frac{(n+p-2)!}{n!}e^{-\nu_n v} \ .
\end{equation}
Substituting the approximate expression for $\nu_n$ into the summation above,
we get after some intermediate steps the following result:
\begin{equation}
S=\frac{(p-1)!}{2^{p-1}}\frac{\cosh(v/2)}{\sinh^{p}(v/2)}\left[1-\frac{vp\Delta^2}{2\sinh(v)}\left(1+4\xi\sinh^2(v/2)\right)\right] \ .
\end{equation}
Substituting this expression into (\ref{Wfunc}) we obtain (\ref{Wight2}).

\section{Hadamard function in five-dimensions}
\label{Ap:B}
Following the procedure given in \cite{Chris2}, in this appendix we shall briefly present the Hadamard function for a five-dimensional spacetime. In this way we start with the expression
\begin{equation}
\label{HD5}
G(x,x')=\frac{i\pi\Delta^{1/2}(x,x')}{(4\pi i)^{5/2}}\sum_{k=0}^\infty a_k(x,x')\left(-\frac\partial{\partial m^2}\right)^k\left(-\frac{z} {2im^2}\right)^{-3/2}H_{3/2}^{(2)}(z) \ ,
\end{equation}
where, $z^2=-2m^2\sigma$, $\Delta(x,x')$ is the Van Vleck-Morette determinant, and $H_\mu^{(2)}$ the Hankel function, given by \cite{Grad}
\begin{equation}
H_{3/2}^{(2)}(z)=-\sqrt{\frac2{\pi z}}e^{-iz}\left(1-\frac iz\right) \ .
\end{equation}
Substituting the above expression into (\ref{HD5}) we get,
\begin{equation}
G(x,x')=\frac{i\Delta^{1/2}(x,x')}{8\pi^2}\frac1{2\sigma}\sum_{k=0}^\infty a_k(x,x')(-1)^k\frac{\partial^k}{\partial (m^2)^k}(m^2)^{1/2} \ e^{\sqrt{2\sigma m^2}} \left(1-\frac1{\sqrt{2\sigma m^2}}\right) \ .
\end{equation}
The expansion, up to the term $a_1$, reproduces (\ref{GH5}).

\end{document}